\documentclass[preprint,12pt]{elsarticle}%
\usepackage{amsfonts}
\usepackage{amsmath}%
\usepackage{amssymb}%
\usepackage{graphicx}
\providecommand{\U}[1]{\protect\rule{.1in}{.1in}}
\journal{ArXiv}
\newtheorem{theorem}{Theorem}

\newtheorem{example}[theorem]{Example}

\newtheorem{lemma}[theorem]{Lemma}

\newtheorem{proposition}[theorem]{Proposition}
\newtheorem{remark}[theorem]{Remark}

\begin{document}
%
\begin{frontmatter}%


%

\title{Similarity of general population matrices and pseudo-Leslie matrices}%

%

\author{Jo\~ao F. Alves\footnote{
jalves@math.tecnico.ulisboa.pt} and Henrique M. Oliveira\footnote
{corresponding:
holiv@math.tecnico.ulisboa.pt}}%
%

\address{Center of Mathematical Analysis, Geometry and Dynamical
Sistems, Departamento de Matem\'atica, Instituto Superior T\'ecnico,
Universidade de Lisboa, Av. Rovisco Pais, 1049-001 Lisbon, Portugal\\}%
%

\begin{abstract}%

A similarity transformation is obtained between general population matrices
models of the Usher or Lefkovitch types and a simpler model, the pseudo-Leslie
model. The pseudo Leslie model is a matrix that can be decomposed in a row
matrix, which is not necessarily non-negative and a subdiagonal positive
matrix. This technique has computational advantages, since the solutions of
the iterative problem using Leslie matrices are readily obtained . In the case
of two age structured population models, one Lefkovitch and another Leslie,
the Kolmogorov-Sinai entropies are different, despite the same growth ratio of
both models. We prove that Markov matrices associated to similar population
matrices are similar.%

\end{abstract}%
%

%

\end{frontmatter}%



\section{Introduction}

This article deals with classic discrete structured models for linear
population dynamics \cite{Cushing,Pollard} such as Leslie matrices and
Lefkovitch or Usher matrices. Giving $A$, a non negative $n\times n$ matrix
and a population vector $\mathbf{x}_{k}$ which components are the fractions of
the population at each age or stage, the dynamical system that gives the
population vector at any positive time $k+1$ is given by%
\[
\mathbf{x}_{k+1}=A\mathbf{x}_{k}\text{, with initial condition }\mathbf{x}%
_{0}\text{.}%
\]
Obviously the solution is given by the powers of $A$%
\[
\mathbf{x}_{k}=A^{k}\mathbf{x}_{0}\text{.}%
\]

In this paper we prove that there is a similarity transform that converts the
complicated dynamics of the so called Usher or Lefkovitch matrices to the
simpler study of matrices which are Leslie matrices or pseudo-Leslie matrices,
a concept that we introduce in this paper.

The paper is organized in three sections, in the second we introduce
pseudo-Leslie matrices and prove the main theorem. In the third section we
present some consequences of interest in population dynamics, namely on the
similarity of Markov matrices associated to similar population dynamics
matrices and obtain transformation rules for corresponding stationary distributions.

\section{Main theorem}

In age structured population dynamics one divides the population in classes
\cite{Cushing,Logofet}. When we consider size classes or stage classes instead
of pure age classes we have a structured population model with dynamics given
by the linear equation%
\begin{equation}
\mathbf{x}_{n+1}=\mathcal{L}\mathbf{x}_{n}, \label{SystDyna}%
\end{equation}
where $\mathbf{x}_{n}$ is a non negative structured absolute population
vector, or a proportion of individuals in each class and $\mathcal{L}$ is a
matrix such that%
\[
\mathcal{L}=\left[
\begin{array}
[c]{cccccc}%
f_{1} & f_{2} & f_{3} & \cdots & f_{n-1} & f_{n}\\
b_{1} & c_{1} & 0 & \cdots & 0 & 0\\
0 & b_{2} & c_{2} & \cdots & 0 & 0\\
\vdots & \vdots & \vdots & \ddots & \vdots & \vdots\\
0 & 0 & 0 & \cdots & c_{n-2} & 0\\
0 & 0 & 0 & \cdots & b_{n-1} & c_{n-1}%
\end{array}
\right]  ,
\]
usually called Usher (in the classic reference \cite{Cushing}) or Lefkovitch
matrix in \cite{Logofet}. The coefficient $f_{j}$ is called the fertility rate
of class $j>1$, the coefficient $b_{k}>0,$ for any $k=1,\ldots,n-1$, is the
transition rate from class $k-1$ to class $k$ and the $c_{l}$ the rate of
individuals that remain in class $l$. Along this paper we assume that
$f_{n}>0$, assuring that $\mathcal{L}$\ is irreducible \cite{Cushing}.

The coefficient $f_{1}$ can be decomposed in $\widehat{f}_{1}+c_{0}$, i.e., a
fertility rate and a permanency rate. Since this decomposition has no
influence on the similarity transformation we do not split $f_{1}$. One must
keep in mind the biological meaning of this coefficient.

The solution of the problem is given by the powers of $\mathcal{L}$, given the
non-negative initial condition $\mathbf{x}_{0}$%
\[
\mathbf{x}_{n}=\mathcal{L}^{n}\mathbf{x}_{0}\text{.}%
\]

A Leslie matrix is a matrix of the type
\[
L=\left[
\begin{array}
[c]{ccccc}%
\phi_{1} & \phi_{2} & \cdots & \phi_{n-1} & \phi_{n}\\
b_{1} & 0 & \cdots & 0 & 0\\
0 & b_{2} & \cdots & 0 & 0\\
\vdots & \vdots & \ddots & \vdots & \vdots\\
0 & 0 & \cdots & b_{n-1} & 0
\end{array}
\right]  ,
\]
where all the entries $\phi_{j}$ are non-negative and all $b_{j}$ are strictly
positive. The Leslie matrix can be decomposed in two matrices%
\[
L=R+B,
\]
where%
\[
R=\left[
\begin{array}
[c]{ccccc}%
\phi_{1} & \phi_{2} & \cdots & \phi_{n-1} & \phi_{n}\\
0 & 0 & \cdots & 0 & 0\\
0 & 0 & \cdots & 0 & 0\\
\vdots & \vdots & \ddots & \vdots & \vdots\\
0 & 0 & \cdots & 0 & 0
\end{array}
\right]  \text{ and }B=\left[
\begin{array}
[c]{ccccc}%
0 & 0 & \cdots & 0 & 0\\
b_{1} & 0 & \cdots & 0 & 0\\
0 & b_{2} & \cdots & 0 & 0\\
\vdots & \vdots & \ddots & \vdots & \vdots\\
0 & 0 & \cdots & b_{n-1} & 0
\end{array}
\right]  \text{.}%
\]
When the entries $\phi_{n}$ of the first row of $R$ are real numbers, not
restricted to the non-negative case, we say that $L$ is a pseudo-Leslie
matrix. Obviously this class of matrix does not have an immediate biological
correspondence when some of its entries are negative. That poses no problem in
the framework of this article, since $L$ is merely used as a computational instrument.

To state the main theorem we define the sums of products of $p$ factors
$\Gamma_{i}^{p}$, where $i=1,...,n$ denotes the row index of a given $n\times
n$ Lefkovitch matrix $\mathcal{L}$
\[
\Gamma_{i}^{p}=
\]%
\[
\left\{
\begin{array}
[c]{ll}%
\left(  -1\right)  ^{p}%
{\displaystyle\sum\limits_{n-1\geq i_{p}>\cdots>i_{2}>i_{1}\geq i}}
c_{i_{1}}c_{i_{2}}\cdots c_{i_{p}} & \text{if }0<p\leq n-i\\
1 & \text{if }p=0\\
0 & \text{if }n-i<p
\end{array}
\right.  .
\]
For the products of the transition rates $b_{1},...,b_{n-1}$ of $\mathcal{L}$
we use the notation%
\[
\Lambda_{i}^{j}=\left\{
\begin{array}
[c]{ll}%
{\displaystyle\prod\limits_{k=i}^{j}}
b_{k}\medskip & \text{if }i\leq j\leq n-1\\
1 & \text{if }j=i-1
\end{array}
\right.  .
\]
Now we introduce an upper triangular matrix $S$ and a pseudo-Leslie matrix $L$
defined by%
\[
S=\left[
\begin{array}
[c]{cccccc}%
1 & s_{1,2} & s_{1,3} & \cdots & s_{1,n-1} & s_{1,n}\\
0 & 1 & s_{2,3} & \cdots & s_{2,n-1} & s_{2,n}\\
0 & 0 & 1 & \cdots & s_{3,n-1} & s_{3,n}\\
\vdots & \vdots & \vdots & \ddots & \vdots & \vdots\\
0 & 0 & 0 & \cdots & 1 & s_{n-1,n}\\
0 & 0 & 0 & \cdots & 0 & 1
\end{array}
\right]  \text{,}%
\]
with%
\[
s_{i,j}=\frac{\Gamma_{i}^{j-i}}{\Lambda_{i}^{j-1}}\text{, for }j\geq i
\]
and%
\[
L=\left[
\begin{array}
[c]{cccccc}%
\phi_{1} & \phi_{2} & \phi_{3} & \cdots & \phi_{n-1} & \phi_{n}\\
b_{1} & 0 & 0 & \cdots & 0 & 0\\
0 & b_{2} & 0 & \cdots & 0 & 0\\
\vdots & \vdots & \vdots & \ddots & \vdots & \vdots\\
0 & 0 & 0 & \cdots & 0 & 0\\
0 & 0 & 0 & \cdots & b_{n-1} & 0
\end{array}
\right]  ,
\]
with%
\[
\phi_{j}=-\frac{\Gamma_{1}^{j}}{\Lambda_{1}^{j-1}}+%
{\displaystyle\sum\limits_{k=1}^{j}}
\frac{\Gamma_{k}^{j-k}}{\Lambda_{k}^{j-1}}f_{k}\text{, for }j=1,...,n.
\]
We are now in position to state the main result of this work.

\begin{theorem}
\label{thm}For any Lefkovitch matrix, $\mathcal{L}$, one has $S^{-1}%
\mathcal{L}S=L$ where $S$ and $L$ are the matrices defined above.
\end{theorem}

The following lemma is used in the proof of theorem \ref{thm}.

\begin{lemma}
\label{lem}If $\mathcal{L}$ is a $n\times n$ Lefkovitch matrix, then
$\Gamma_{i}^{p+1}=c_{i-1}\Gamma_{i}^{p}+\Gamma_{i-1}^{p+1}$, for all $p\geq0$
and $n\geq i>1$.
\end{lemma}

\smallskip\noindent\textbf{Proof.} As $\Gamma_{i}^{0}=1$, $\Gamma_{i}%
^{1}-\Gamma_{i-1}^{1}=c_{i-1}$ and $\Gamma_{i}^{p}=\Gamma_{i}^{p+1}%
=\Gamma_{i-1}^{p+1}=0$ for $p>n-i$, the proof is obvious for $p=0$ or $p>n-i$
. So we may assume $0<p\leq n-i$.

If $0<p<n-i$, then%
\begin{align*}
\Gamma_{i-1}^{p+1}  &  =\left(  -1\right)  ^{p+1}%
{\displaystyle\sum\limits_{n-1\geq i_{p+1}>\cdots>i_{2}>i_{1}\geq i-1}}
c_{i_{1}}c_{i_{2}}\ldots c_{i_{p+1}}\\
&  =\left(  -1\right)  ^{p+1}%
{\displaystyle\sum\limits_{n-1\geq i_{p+1}>\cdots>i_{2}>i_{1}\geq i}}
c_{i_{1}}c_{i_{2}}\ldots c_{i_{p+1}}\\
&  -\left(  -1\right)  ^{p}%
{\displaystyle\sum\limits_{n-1\geq i_{p+1}>\cdots>i_{2}\geq i}}
c_{i-1}c_{i_{2}}\ldots c_{i_{p+1}}\\
&  =\Gamma_{i}^{p+1}-c_{i-1}\Gamma_{i}^{p}.
\end{align*}
Finally, assume that $0<p=n-i$. In this case, as $\Gamma_{i}^{p+1}=0$, one
gets%
\[
c_{i-1}\Gamma_{i}^{p}+\Gamma_{i-1}^{p+1}=
\]
\begin{align*}
&  =c_{i-1}\left(  -1\right)  ^{p}c_{i}c_{i+1}\ldots c_{n-1}+\left(
-1\right)  ^{p+1}c_{i-1}c_{i}\ldots c_{n-1}\\
&  =0=\Gamma_{i}^{p+1}\text{. }\square
\end{align*}

We are now in position to prove the main result.

\smallskip\noindent\textbf{Proof of theorem \ref{thm}.} In order to prove the
equality $\mathcal{L}S=SL$, we begin by computing $SL$. As $s_{i,i}=1$ and
$s_{i,j}=0$ for $i>j$, one has
\begin{align*}
\left(  SL\right)  _{i,j}  &  =\left\{
\begin{array}
[c]{ll}%
s_{i,1}\phi_{n} & \text{if }j=n\\
s_{i,1}\phi_{j}+s_{i,j+1}b_{j} & \text{if }j<n
\end{array}
\right. \\
&  =\left\{
\begin{array}
[c]{ll}%
\phi_{n} & \text{if }i=1\text{, }j=n\text{ }\\
\phi_{j}+s_{1,j+1}b_{j} & \text{if }i=1\text{, }j<n\\
b_{j} & \text{if }i=j+1\\
s_{i,j+1}b_{j} & \text{if }n>j\geq i>1\\
0 & \text{otherwise}%
\end{array}
\right.  \text{.}%
\end{align*}
As $\Gamma_{1}^{n}=0$ and $\Lambda_{1}^{j}=\Lambda_{1}^{j-1}b_{j}$ one has%
\begin{align*}
\phi_{n}  &  =-\frac{\Gamma_{1}^{n}}{\Lambda_{1}^{n-1}}+%
{\displaystyle\sum\limits_{k=1}^{n}}
\frac{\Gamma_{k}^{n-k}}{\Lambda_{k}^{n-1}}f_{k}\\
&  =%
{\displaystyle\sum\limits_{k=1}^{n}}
\frac{\Gamma_{k}^{n-k}}{\Lambda_{k}^{n-1}}f_{k}\text{,}%
\end{align*}
and%
\begin{align*}
\phi_{j}+s_{1,j+1}b_{j}  &  =-\frac{\Gamma_{1}^{j}}{\Lambda_{1}^{j-1}}+%
{\displaystyle\sum\limits_{k=1}^{j}}
\frac{\Gamma_{k}^{j-k}}{\Lambda_{k}^{j-1}}f_{k}+\frac{\Gamma_{1}^{j}}%
{\Lambda_{1}^{j}}b_{j}\\
&  =-\frac{\Gamma_{1}^{j}}{\Lambda_{1}^{j-1}}+%
{\displaystyle\sum\limits_{k=1}^{j}}
\frac{\Gamma_{k}^{j-k}}{\Lambda_{k}^{j-1}}f_{k}+\frac{\Gamma_{1}^{j}}%
{\Lambda_{1}^{j-1}}\\
&  =%
{\displaystyle\sum\limits_{k=1}^{j}}
\frac{\Gamma_{k}^{j-k}}{\Lambda_{k}^{j-1}}f_{k},\text{ for }j<n\text{,}%
\end{align*}
finally we get%
\[
s_{i,j+1}b_{j}=\frac{\Gamma_{i}^{j+1-i}}{\Lambda_{i}^{j}}b_{j}=\frac
{\Gamma_{i}^{j+1-i}}{\Lambda_{i}^{j-1}}\text{, for }n>j\geq i\text{.}%
\]
Thus, we may write%
\[
\left(  SL\right)  _{i,j}=\left\{
\begin{array}
[c]{ll}%
{\displaystyle\sum\limits_{k=1}^{j}}
\frac{\Gamma_{k}^{j-k}}{\Lambda_{k}^{j-1}}f_{k} & \text{if }i=1\\
b_{j} & \text{if }i=j+1\\
\frac{\Gamma_{i}^{j+1-i}}{\Lambda_{i}^{j-1}} & \text{if }n>j\geq i>1\\
0 & \text{otherwise}%
\end{array}
\right.  .
\]
Notice that since $\Gamma_{i}^{n+1-i}=0$ for all $i$, we finally arrive at%
\begin{equation}
\left(  SL\right)  _{i,j}=\left\{
\begin{array}
[c]{ll}%
{\displaystyle\sum\limits_{k=1}^{j}}
\frac{\Gamma_{k}^{j-k}}{\Lambda_{k}^{j-1}}f_{k} & \text{if }i=1\\
b_{j} & \text{if }i=j+1\\
\frac{\Gamma_{i}^{j+1-i}}{\Lambda_{i}^{j-1}} & \text{if }j\geq i>1\\
0 & \text{otherwise}%
\end{array}
\right.  . \label{f1}%
\end{equation}
Next we compute $\mathcal{L}S$. As $s_{i,i}=1$ and $s_{i,j}=0$ for $i>j$, one
has%
\begin{align*}
\left(  \mathcal{L}S\right)  _{i,j}  &  =\left\{
\begin{array}
[c]{cc}%
\sum_{k=1}^{n}s_{k,j}f_{k} & \text{if }i=1\\
b_{i-1}s_{i-1,j}+c_{i-1}s_{i,j} & \text{if }i>1
\end{array}
\right. \\
&  =\left\{
\begin{array}
[c]{ll}%
\sum_{k=1}^{j}s_{k,j}f_{k} & \text{if }i=1\\
b_{j} & \text{if }i=j+1\\
b_{i-1}s_{i-1,j}+c_{i-1}s_{i,j} & \text{if }j\geq i>1\\
0 & \text{otherwise}%
\end{array}
\right. \\
&  =\left\{
\begin{array}
[c]{ll}%
\sum_{k=1}^{j}\frac{\Gamma_{k}^{j-k}}{\Lambda_{k}^{j-1}}f_{k} & \text{if
}i=1\\
b_{j} & \text{if }i=j+1\\
b_{i-1}\frac{\Gamma_{i-1}^{j-i+1}}{\Lambda_{i-1}^{j-1}}+c_{i-1}\frac
{\Gamma_{i}^{j-i}}{\Lambda_{i}^{j-1}} & \text{if }j\geq i>1\\
0 & \text{otherwise}%
\end{array}
\right.  .
\end{align*}
As $\Lambda_{i-1}^{j-1}=b_{i-1}\Lambda_{i}^{j-1}$ for $j\geq i>1$, one has%
\begin{align*}
b_{i-1}\frac{\Gamma_{i-1}^{j-i+1}}{\Lambda_{i-1}^{j-1}}+c_{i-1}\frac
{\Gamma_{i}^{j-i}}{\Lambda_{i}^{j-1}}  &  =\frac{\Gamma_{i-1}^{j-i+1}}%
{\Lambda_{i}^{j-1}}+c_{i-1}\frac{\Gamma_{i}^{j-i}}{\Lambda_{i}^{j-1}}\\
&  =\frac{\Gamma_{i-1}^{j-i+1}+c_{i-1}\Gamma_{i}^{j-i}}{\Lambda_{i}^{j-1}}%
\end{align*}
and consequently
\begin{equation}
\left(  \mathcal{L}S\right)  _{i,j}=\left\{
\begin{array}
[c]{ll}%
\sum_{k=1}^{j}\frac{\Gamma_{k}^{j-k}}{\Lambda_{k}^{j-1}}f_{k} & \text{if
}i=1\\
b_{j} & \text{if }i=j+1\\
\frac{\Gamma_{i-1}^{j-i+1}+c_{i-1}\Gamma_{i}^{j-i}}{\Lambda_{i}^{j-1}} &
\text{if }j\geq i>1\\
0 & \text{otherwise}%
\end{array}
\right.  . \label{f2}%
\end{equation}
Now, using lemma \ref{lem} we see that (\ref{f1}) and (\ref{f2}) are the same,
which completes the proof. $\square$

The dynamical system (\ref{SystDyna}) can be solved using the easily
computable powers of $L$%
\[
\mathbf{x}_{n}=\mathcal{L}^{n}\mathbf{x}_{0}=S^{-1}L^{n}S\mathbf{x}%
_{0}\text{.}%
\]
Since $\mathcal{L}$ and $L$ are similar, they share the same spectrum and the
Perron-Frobenius Theorem still holds for $L$ in what concerns the existence of
a simple dominant positive eigenvalue. Using a generating function and formal
power series obtained in \cite{Alves} or the classic Jordan canonical form, it
is always possible to obtain the powers of $L$. The eigenvectors of
$\mathcal{L}$ will be studied in the next section.

\section{Sinai Kolmogorov entropy, Markov matrices and stationary
distributions}

In this section, using a simple example, we show that the Kolmogorov-Sinai
entropy \cite{Dem1,Dem5,Dem7,Dem8} is not an algebraic invariant. We also
establish that two Markov matrices associated \cite{Dem8} to population
dynamics similar matrices\footnote{Under very general conditions.} are
similar. Finally, we establish a transformation rule for the two stationary
distributions of Markov matrices associated with two similar population matrices.

Given two matrices, one of Lefkovitch type and the other of Leslie
type\footnote{We consider a true non-negative Leslie matrix to establish this
conclusion.}, with the same growth rate, they can have different
Sinai-Kolmogorov entropies as we see in the following example.

\begin{example}
Let
\[
\mathcal{L}=\left[
\begin{array}
[c]{cc}%
1 & 3\\
0.4 & 0.55
\end{array}
\right]  \text{,}%
\]
we have the similarity matrix%
\[
S=\left[
\begin{array}
[c]{cc}%
1 & -1.375\\
0 & 1
\end{array}
\right]  \text{,}%
\]
and a Leslie matrix $L$ similar to $\mathcal{L}$, which is%
\[
L=\left[
\begin{array}
[c]{cc}%
1.55 & 1.625\\
0.4 & 0
\end{array}
\right]  \text{.}%
\]
The Perron-Frobenius dominant eigenvalue is $\lambda=1.89331$ both for $L$ and
$\mathcal{L}$.
The Markov matrix $P^{A}$ \cite{Dem8}, corresponding to a population matrix $A$ is
obtained using the relations%
\[
p_{ij}^{A}=\frac{a_{ij}\text{ }u_{j}}{\lambda u_{i}}\text{,}%
\]
where $\lambda$ is the dominant eigenvalue of $A$, and the column vector
$\mathbf{u}=\left(  u_{i}\right)  _{i=1,\ldots,n}>0$ is the Perron-Frobenius
right eigenvector of $A$. (The left eigenvector will be called the line vector
$\mathbf{v}=\left(  v_{i}\right)  _{i=1,\ldots,n}^{T}$). For the Lefkovitch matrix
$\mathcal{L}$ we get the associated Markov matrix%
\[
P^{\mathcal{L}}=\left[
\begin{array}
[c]{cc}%
0.528175 & 0.471825\\
0.709504 & 0.290496
\end{array}
\right]  \text{,}%
\]
the stationary distribution of $P^{\mathcal{L}}$ is $\mathbf{\pi}%
^{\mathcal{L}}=\left[
\begin{array}
[c]{cc}%
0.600598 & 0.399402
\end{array}
\right]  $. The population Sinai-Kolmogorov entropy \cite{Dem8} is%
\[
H_{\mathcal{L}}=-\sum_{i,j}^{2}\pi_{i}^{\mathcal{L}}p_{ij}^{\mathcal{L}}\log
p_{ij}^{\mathcal{L}}\text{,}%
\]
where $p_{ij}^{\mathcal{L}}$ are the entries of $P^{\mathcal{L}}$ and $\pi
_{i}^{\mathcal{L}}$ are the components of the stationary distribution
$\mathbf{\pi}^{\mathcal{L}}$ of $P^{\mathcal{L}}$ (the left eigenvector
associated with the Perron-Frobenius eigenvalue $1$ of $P^{\mathcal{L}}$, such
that $\mathbf{\pi}^{\mathcal{L}}P^{\mathcal{L}}=\mathbf{\pi}^{\mathcal{L}}$).
Doing the same computation for $L$ we have%
\[
H_{L}=-\sum_{i,j}^{2}\pi_{i}^{L}p_{ij}^{L}\log p_{ij}^{L}\text{,}%
\]
where $P^{L}$ is the matrix with entries $p_{ij}^{L}$, the Markov matrix
associated to $L$ is%
\[
P^{L}=\left[
\begin{array}
[c]{cc}%
0.818671 & 0.181329\\
1 & 0
\end{array}
\right]  \text{.}%
\]
The stationary distribution of $P^{L}$ is $\mathbf{\pi}^{L}=\left[
\begin{array}
[c]{cc}%
0.846504 & 0.153496
\end{array}
\right]  $ and the entropies of $\mathcal{L}$ and $L$ are different,
respectively $H_{\mathcal{L}}=0.656027$ and $H_{L}=0.400738$.
\end{example}

The Markov matrices $P^{L}$ and $P^{\mathcal{L}}$ associated to $L$ and
$\mathcal{L}$\ are also similar, with the same eigenvalues as we will see
below. This result can be stated in the general context of similar
matrices\footnote{Not necessarily Lefkovitch, Usher or Leslie matrices.} under
the following hypothesis, which are assumed until the end of the paper:

\begin{enumerate}
\item \label{c1}$\mathcal{L}$ is non-negative and irreducible, therefore has
the dominant eigenvalue $\lambda$, and associated left and right positive
eigenvectors $\mathbf{t}$ and $\mathbf{w}$, respectively.

\item \label{c2}$L$ and $\mathcal{L}$ are similar, related by the invertible
similarity matrix $S,$ such that $\mathcal{L}S=SL$.

\item \label{c3}$L$, not necessarily non-negative, has right and left
eigenvectors, respectively $\mathbf{u}$ and $\mathbf{v}$, associated to
$\lambda$ with all entries positive.
\end{enumerate}

The right eigenvector of $L$ associated to the dominant eigenvalue $\lambda$
\[
L\mathbf{u}=\lambda\mathbf{u}%
\]
is related to the right eigenvector $\mathbf{w}$ of $\mathcal{L}$ by the
transformation rule $\mathbf{w=}S\mathbf{u}$, since%
\[
\mathcal{L}S\mathbf{u=}\lambda S\mathbf{u\Leftrightarrow}\mathcal{L}%
\mathbf{w=}\lambda\mathbf{w}\text{.}%
\]
The same happens for the left eigenvector $\mathbf{v}$ of $L$
\[
\mathbf{v}L=\lambda\mathbf{v}%
\]
and the left eigenvector $\mathbf{t=v}S^{-1}$ of $\mathcal{L}$, since%
\[
\mathbf{v}S^{-1}\mathcal{L}\mathbf{=}\lambda\mathbf{v}S^{-1}%
\mathbf{\Leftrightarrow t}\mathcal{L}\mathbf{=}\lambda\mathbf{t}\text{.}%
\]
The Markov matrix associated with $L$ \cite{Dem8} is given by its entries%
\[
p_{ij}^{L}=\frac{L_{ij}\text{ }u_{j}}{\lambda u_{i}}\text{.}%
\]
On the other hand, the Markov matrix associated with $\mathcal{L}$ is given by%
\[
p_{ij}^{\mathcal{L}}=\frac{\mathcal{L}_{ij}\text{ }w_{j}}{\lambda w_{i}%
}\text{.}%
\]
The stationary distribution \cite{Dem8} of $P^{L}$ is%
\[
\mathbf{\pi}^{L}=\frac{\left[
\begin{array}
[c]{cccc}%
v_{1}u_{1} & v_{2}u_{2} & \ldots & v_{n}u_{n}%
\end{array}
\right]  }{\mathbf{vu}}\text{,}%
\]
where $\mathbf{vu}$ is a compact notation for the inner product of the line
vector $\mathbf{v}$ and the column vector $\mathbf{u}$. The stationary
distribution of $\mathcal{L}$ is%
\[
\mathbf{\pi}^{\mathcal{L}}=\frac{\left[
\begin{array}
[c]{cccc}%
t_{1}w_{1} & t_{2}w_{2} & \ldots & t_{n}w_{n}%
\end{array}
\right]  }{\mathbf{tw}}\text{.}%
\]
It is possible to prove that the Markov matrices $P^{L}$ and $P^{\mathcal{L}}
$ are similar.

\begin{proposition}
$P^{L}$ and $P^{\mathcal{L}}$ are similar if $L$ and $\mathcal{L}$ are similar.
\end{proposition}

\smallskip\noindent\textbf{Proof.} One defines the square matrices $U$ and $W
$ such that
\[
U=\left[
\begin{array}
[c]{cccc}%
u_{1} &  &  & \\
& u_{2} &  & \\
&  & \ddots & \\
&  &  & u_{n}%
\end{array}
\right]  ,\text{ }W=\left[
\begin{array}
[c]{cccc}%
w_{1} &  &  & \\
& w_{2} &  & \\
&  & \ddots & \\
&  &  & w_{n}%
\end{array}
\right]
\]
with all $u_{i}\neq0$ and $w_{i}\neq0$, the inverses of $U$ and $W$ are%
\[
U^{-1}=\left[
\begin{array}
[c]{cccc}%
\frac{1}{u_{1}} &  &  & \\
& \frac{1}{u_{2}} &  & \\
&  & \ddots & \\
&  &  & \frac{1}{u_{n}}%
\end{array}
\right]  \text{, }W^{-1}=\left[
\begin{array}
[c]{cccc}%
\frac{1}{w_{1}} &  &  & \\
& \frac{1}{w_{2}} &  & \\
&  & \ddots & \\
&  &  & \frac{1}{w_{n}}%
\end{array}
\right]  \text{.}%
\]
With this notation consider the transformations%
\[
P^{L}=\frac{1}{\lambda}U^{-1}LU\text{ and }P^{\mathcal{L}}=\frac{1}{\lambda
}W^{-1}\mathcal{L}W\text{,}%
\]
where $\lambda\not =0$.

Now, it is straightforward to prove that $P^{L}$ and $P^{\mathcal{L}}$ are
similar%
\[
P^{\mathcal{L}}=\frac{1}{\lambda}W^{-1}\mathcal{L}W=\frac{1}{\lambda}%
W^{-1}SLS^{-1}W\text{.}%
\]
On the other hand%
\[
P^{L}=\frac{1}{\lambda}U^{-1}LU\text{.}%
\]
Therefore, $\lambda Q$ and $\lambda P$ are similar, since both are similar to
$L$. Explicitly%
\[
L=\lambda S^{-1}WP^{\mathcal{L}}W^{-1}S=\lambda UP^{L}U^{-1}%
\]
or%
\begin{equation}
P^{L}=U^{-1}S^{-1}WP^{\mathcal{L}}W^{-1}SU\text{,} \label{SimilarP}%
\end{equation}
as desired. $\square$

We can prove that $\mathbf{\pi}^{\mathcal{L}}$ is a stationary distribution of
$P^{\mathcal{L}}$ \cite{Dem8} using matrix notation.

\begin{proposition}
The row vector $\mathbf{\pi}^{\mathcal{L}}$ is a stationary distribution of
$P^{\mathcal{L}}$.
\end{proposition}

\smallskip\noindent\textbf{Proof.} Using the left eigenvector $\mathbf{t=}%
\left[
\begin{array}
[c]{cccc}%
t_{1} & t_{2} & \ldots & t_{n}%
\end{array}
\right]  $ of $\mathcal{L}$, we define a diagonal matrix
\[
T=\left[
\begin{array}
[c]{cccc}%
t_{1} &  &  & \\
& t_{2} &  & \\
&  & \ddots & \\
&  &  & t_{n}%
\end{array}
\right]  \text{.}%
\]
We have%
\begin{align*}
\mathbf{\pi}^{\mathcal{L}}P^{\mathcal{L}}  &  =\frac{1}{\lambda\mathbf{tw}%
}\left[
\begin{array}
[c]{cccc}%
1 & 1 & ... & 1
\end{array}
\right]  TWW^{-1}\mathcal{L}W\\
&  =\frac{1}{\lambda\mathbf{tw}}\left[
\begin{array}
[c]{cccc}%
t_{1} & t_{2} & ... & t_{n}%
\end{array}
\right]  \mathcal{L}W\text{,}%
\end{align*}
since $\mathbf{t}$ is a left eigenvector of $\mathcal{L}$ we have%
\begin{align*}
\mathbf{\pi}^{\mathcal{L}}P^{\mathcal{L}}  &  =\frac{1}{\lambda\mathbf{tw}%
}\lambda\left[
\begin{array}
[c]{cccc}%
t_{1} & t_{2} & ... & t_{n}%
\end{array}
\right]  W\\
&  =\frac{TW}{\mathbf{tw}}\\
&  =\mathbf{\pi}^{\mathcal{L}}\text{. }\square
\end{align*}

Using analogous techniques we obtain the relation between the two stationary
distributions of $P^{L}$ and $P^{\mathcal{L}}$.

\begin{proposition}
The stationary distributions $\mathbf{\pi}^{\mathcal{L}}$ and $\mathbf{\pi
}^{L}$ are related by%
\[
\mathbf{\pi}^{L}=\mathbf{\pi}^{\mathcal{L}}W^{-1}SU
\]
\end{proposition}

\smallskip\noindent\textbf{Proof.} From (\ref{SimilarP}) we have%
\[
P^{L}=Z^{-1}P^{\mathcal{L}}Z\text{,}%
\]
where $Z=W^{-1}SU$. In that case the stationary distribution $\mathbf{\pi}%
^{L}$ is given by the relationship%
\[
\mathbf{\pi}^{L}P^{L}=\mathbf{\pi}^{L},
\]
so%
\[
\mathbf{\pi}^{L}Z^{-1}P^{\mathcal{L}}Z=\mathbf{\pi}^{L}\Longleftrightarrow
\mathbf{\pi}^{L}Z^{-1}P^{\mathcal{L}}=\mathbf{\pi}^{L}Z^{-1},
\]
which means that%
\[
\mathbf{\pi}^{\mathcal{L}}=\mathbf{\pi}^{L}Z^{-1}\text{,}%
\]
as desired. $\square$

\begin{remark}
All the results in this section apply to the case of an irreducible Lefkovitch
matrix $\mathcal{L}$ and a similar pseudo-Leslie matrix $L$, since any matrix
of the form%
\[
L=\left[
\begin{array}
[c]{ccccc}%
\phi_{1} & \phi_{2} & \cdots & \phi_{n-1} & \phi_{n}\\
b_{1} & 0 & \cdots & 0 & 0\\
0 & b_{2} & \cdots & 0 & 0\\
\vdots & \vdots & \ddots & \vdots & \vdots\\
0 & 0 & \cdots & b_{n-1} & 0
\end{array}
\right]  ,
\]
with positive coefficients $b_{j}$ and with the dominant eigenvalue $\lambda$
has the positive right eigenvector%
\[
\mathbf{u}=\left[
\begin{array}
[c]{c}%
\Lambda_{1}^{0}\\
\frac{\Lambda_{1}^{1}}{\lambda}\\
\frac{\Lambda_{1}^{2}}{\lambda^{2}}\\
\vdots\\
\frac{\Lambda_{1}^{n-1}}{\lambda^{n-1}}%
\end{array}
\right]  =\left[
\begin{array}
[c]{c}%
1\\
\frac{b_{1}}{\lambda}\\
\frac{b_{1}b_{2}}{\lambda^{2}}\\
\vdots\\
\frac{b_{1}b_{2}\cdots b_{n-1}}{\lambda^{n-1}}%
\end{array}
\right]  \text{.}%
\]
The similar Lefkovitch matrix
\[
\mathcal{L}=\left[
\begin{array}
[c]{cccccc}%
f_{1} & f_{2} & f_{3} & \cdots & f_{n-1} & f_{n}\\
b_{1} & c_{1} & 0 & \cdots & 0 & 0\\
0 & b_{2} & c_{2} & \cdots & 0 & 0\\
\vdots & \vdots & \vdots & \ddots & \vdots & \vdots\\
0 & 0 & 0 & \cdots & c_{n-2} & 0\\
0 & 0 & 0 & \cdots & b_{n-1} & c_{n-1}%
\end{array}
\right]
\]
\ is always irreducible if $f_{n}>0$ and all the $b_{j}$ are positive,
\cite{Cushing}. Therefore, similar Lefkovitch and pseudo-Leslie matrices,
$\mathcal{L}$ and $L$, satisfy conditions \ref{c1}, \ref{c2} and \ref{c3}.
\end{remark}

\section*{Acknowledgement}

The authors were partially funded by FCT/Portugal through project PEst-OE/EEI/LA0009/2013.

\end{document}